\documentclass[showpacs,10pt,letterpaper,notitlepage,superscriptaddress]{revtex4-1}

\usepackage{hyperref}
\usepackage{amsmath}
\usepackage{mathtools}
\usepackage{bbm}
\usepackage{bm}
\usepackage{amsfonts}
\usepackage{amssymb}
\usepackage{mathrsfs}
\usepackage{wasysym}
\usepackage{graphicx}
\usepackage[]{subfigure}
\usepackage{filecontents}
\usepackage[dvipsnames]{xcolor}
\usepackage[T1]{fontenc} 

\hypersetup{
    bookmarks=true,         
    pdftoolbar=true,        
    pdfmenubar=true,        
    pdffitwindow=true,     
    pdfstartview={FitH},     
    pdftitle={My title},     
    pdfauthor={author},      
    pdfsubject={Subject},    
    pdfcreator={Creator},    
    pdfproducer={Producer},  
    pdfkeywords={keyword1} 
    pdfnewwindow=true,       
    colorlinks=true ,        
    linkcolor=blue  ,        
    citecolor=blue,      
    filecolor=green,       
    urlcolor=blue            
}

\begin{document}

\title{On the thermodynamics of the Hayward black hole}

\author{Mart\'in Molina}
\email{martin.molina@alumnos.uv.cl}
\affiliation{Instituto de F\'{i}sica y Astronom\'{i}a, Universidad de Valpara\'{i}so,
	Avenida Gran Breta\~{n}a 1111, Valpara\'{i}so, Chile}

\author{J.R. Villanueva}
\email{jose.villanueva@uv.cl}
\affiliation{Instituto de F\'{i}sica y Astronom\'{i}a, Universidad de Valpara\'{i}so,
Avenida Gran Breta\~{n}a 1111, Valpara\'{i}so, Chile}

\date{\today}

\begin{abstract}
In light of the growing interest in the Hayward black hole solution, a detailed study on the corresponding lapse function and its roots is presented. The lapse function is expressed in terms of the classical Schwarzschild radius $r_s$ and the Hayward's parameter $l$. Both of these quantities are used as thermodynamic variables to find related thermodynamic quantities. In this context, the variable $l$ is associated with a canonical conjugate variable $\mathcal{F}_H$, and a free energy $\Xi$. Moreover, a second order phase transition is found to appears at $l\approx0.333\,r_s$.
\bigskip

{\noindent{\textit{keywords}}: Black Holes, Thermodynamic, Hayward black hole}\\

\end{abstract}

\maketitle

\tableofcontents

\section{Introduction}
Black holes are among the most fascinating objects in the universe, both observationally and theoretically, and as the advances in the related fields grow steadily \cite{Akiyama:2019cqa}, we go towards the better understanding of their nature. In this sense, Einstein's gravitational field equations give a privileged framework in studying black hole, in particular, by means of their spherically symmetric vacuum solution proposed by K. Schwarzschild \cite{Schwarzschild:1916uq}. This solution, as a standard black hole background geometry, can be endowed with a cosmological constant \cite{kottler}, and can form more complex black holes in the presence of electric charge \cite{reissner,nordstrom}, angular momentum \cite{Kerr:1963ud}, or both \cite{Newman:1965my, Gibbons:1977mu}. These black hole solutions have in common a curvature singularity at $r = 0$. In contrast, the so-called {\it regular} or {\it non-singular} black holes do not possess any singularities. The latter have been studied earlier by Bardeen \cite{bardeen}, Hayward \cite{Hayward06}, Dymnikova \cite{Dymnikova:1992ux,Dymnikova:2003vt}, Ay\'on-Beato \& Garc\'ia \cite{AyonBeato:1998ub,AyonBeato:1999ec,AyonBeato:1999rg,AyonBeato:2000zs,AyonBeato:2004ih}, and other authors \cite{Mars_1996,Borde:1996df,Mbonye:2005im,Bronnikov:2005gm,Berej:2006cc,Bambi:2013ufa,Balart:2014cga,Balart:2016zrd,sert}. These studies have inspired further investigations related to such black holes, for example regarding the particle geodesics \cite{Stuchlik:2014qja,Abbas:2014oua,Zhao:2017cwk,chiba17}, and the quasi-normal modes  \cite{Fernando,Flachi:2012nv,Lin:2013ofa,Saleh:2018hba}. Additionally, the black holes remnants and their possible non-commutative effects \cite{Mehdipour:2016vxh}, and the absorption of mass-less scalar waves and the corresponding scattering cross section \cite{Huang:2014nka} have also been investigated. The thermodynamics and phase transitions for regular black holes surrounded by quintessence were studied in Refs.~\cite{Saleh18,Rodrigue20}.

In this paper we consider one of these regular solutions, namely, the one proposed by Hayward. In particular, we focus on studying the parameter $l$ that appears in the solution, and figure out its physical implications. The relevance of this parameter to the Planck length ($\ell_p$) and also to other specific values has been dealt with in the literature, although, in principle, this value is {not} {\it priori} scale determined by Hayward \cite{Hayward06}.
However, since this solution can be constructed by coupling general relativity with a non-linear electromagnetic field, as the degenerate configuration of the gravitational field of a non-linear magnetic monopole \cite{AyonBeato:1999ec,AyonBeato:1999rg,AyonBeato:1998ub,Fan:2016hvf,Fan:2016rih}, one could connect it to the magnetic charges, carried by the configuration.

For this reason, in this work we investigate some interesting consequences obtained by leaving this parameter free to vary. In Sect. \ref{ngse} we introduce the Hayward black hole. In particular, the lapse function and it roots are given a detailed study. In Sect. \ref{termasp}, certain thermodynamic aspects are reviewed and analyzed. Finally, in Sect. \ref{Summary}, we conclude with some final remarks.

\section{The Hayward black hole}\label{ngse}

To derive regular spacetime solutions from general relativity, when coupling with a non-linear electromagnetic field is present, we encounter the action \cite{AyonBeato:1998ub,AyonBeato:1999ec,AyonBeato:1999rg,AyonBeato:2000zs}
\begin{equation}
    \label{action}I=\frac{1}{2 \kappa}\int {\rm d}^4x\sqrt{-g} \,\left(R-2\kappa \mathcal{L}(\mathfrak{F})\right),
\end{equation}where $\kappa=8\pi G/c^4$, with $G$ as the Newton gravitational constant and $c$ as the speed of light. Here, we have defined $\mathfrak{F}\equiv F_{\mu \nu}F^{\mu \nu}$, where $\bm{F}=\mathrm{d}\bm{A}$ is the field strength of the electromagnetic field associated with the vector potential $\bm{A}$. Accordingly, $\mathcal{L}$ is the electromagnetic Lagrangian density expressed in terms of $\mathfrak{F}$ \cite{Bronnikov:2000vy,Fan:2016hvf,Fan:2016rih,Bronnikov:2017tnz,Toshmatov:2018cks}. Therefore, the covariant field equations are read as
\begin{equation}
    \label{cfe} G_{\mu \nu} =\kappa\, T_{\mu \nu}, 
\end{equation}in which, $G_{\mu \nu}=R_{\mu \nu}-\frac{1}{2}R\,g_{\mu \nu}$ is the Einstein tensor. {The tensor $F_{\mu \nu}$, in fact, obeys the dynamical equations together with
\begin{equation}
\label{fmneqs} \nabla_{\mu}(\mathcal{L}_{\mathfrak{F}}\,F^{\mu \nu})=0,\qquad \nabla_{\mu}\,\left(\star F^{\mu \nu}\right)=0,
\end{equation}where $\star$ denotes the Hodge dual}, and $\mathcal{L}_{\mathfrak{F}}\equiv \frac{\partial \mathcal{L}}{\partial \mathfrak{F}}$. Hence, the energy-momentum tensor becomes
\begin{equation}
    \label{emten}T_{\mu}^{\nu} = 2\mathcal{L}_{\mathfrak{F}}\,F_{\mu \alpha}\,F^{\nu \alpha}-\frac{1}{2}\,\delta_{\mu}^{\nu}\mathcal{L}.
\end{equation}
{If spherical symmetry is assumed, the tensor $F_{\mu \nu}$ can
involve only a radial electric field $F_{t r}=-F_{r t}$ and a radial magnetic field $F_{\theta \phi}=-F_{\theta \phi}$, so Eqs. (\ref{fmneqs}) become
\begin{equation}
\label{radelmag}r^2 \mathcal{L}_{\mathfrak{F}} F^{t r}=q_e,\quad F_{\theta \phi}=q_m \sin\theta,
\end{equation}where $q_e$ and $q_m$ denote respectively the electric and magnetic charges. Nevertheless, Bronnnikov has shown in Ref.~\cite{Bronnikov:2000vy} that,
when $\mathcal{L}(\mathfrak{F})$ has a Maxwell asymptotic (i.e., when $\mathcal{L}\rightarrow 0, \mathcal{L}_{\mathfrak{F}}\rightarrow 1$ as $\mathfrak{F}\rightarrow 0$), the field system given by Eqs. (\ref{cfe}) and (\ref{emten}) does not admit a static spherically symmetric solution with a regular center and a nonzero electric charge, and thus, the electromagnetic field strength will be
\begin{equation}
\label{stf1a}\mathfrak{F}=2F_{\theta \phi} F^{\theta \phi}=\frac{2 q_m^2}{r^4}.
\end{equation}
 }
{Therefore,} the regular, non-singular, static spherically symmetric black hole spacetime presented by Hayward in Ref.~\cite{Hayward06}, can be obtained from the Lagrangian density \cite{Bronnikov:2000vy,Fan:2016hvf,Fan:2016rih,Bronnikov:2017tnz,Toshmatov:2018cks}
\begin{equation}
    \label{denshayw} \mathcal{L}=\frac{6}{l^2}\,\frac{(2 l^2 \mathfrak{F})^\frac{3}{2}}{[1+(2 l^2 \mathfrak{F})^\frac{3}{4}]^2},
\end{equation}and is described by the following ansatz: \begin{eqnarray}
	\label{eq1}{\rm d}s^2&=&-f(r)\,{\rm d}(c t)^2+\frac{{\rm d}r^2}{f(r)}+r^2{\rm d}\theta^2+r^2 \sin^2\theta \,{\rm d}\phi^2, \\
	\label{amg}A_{\phi}&=&q_m\cos \theta,
\end{eqnarray}
where the lapse function is given by 
\begin{eqnarray}
\label{eq2} &&f(r)=1-\frac{r_s\, r^2}{r^3+r_s\, l^2}=\frac{P_3(r; r_s, l)}{r^3+r_s\, l^2},\\
\label{eq3} &&P_3(r; r_s, l)=r^3-r_s\, r^2+r_s\, l^2.
\end{eqnarray}
Here, $r_s=2 G M/c^2$ is the {\it classical Schwarzschild radius}, $M$ is the gravitational energy--mass of the system, and  $l$ is the Hayward's parameter, whose value, in principle, will be restricted to the range $0\leq l <\infty$. {It should be however noted that, in the his original work, Hayward had considered this parameter to be of the order of Planck's length, and it was then related to a magnetic charge through the definition}
\begin{equation}
    \label{magchaHa}q_m=\frac{\sqrt[3]{r_s^2\,l}}{2},
\end{equation}
following the discussions given in Refs. \cite{Bronnikov:2000vy,Fan:2016hvf,Fan:2016rih,Bronnikov:2017tnz,Toshmatov:2018cks}. Hence, the required condition $q_m>0$ corresponds to a black hole solution with $l>0$. Accordingly, the square of the field strength tensor (\ref{stf1a}) has the value
\begin{equation}
    \label{stf1}\mathfrak{F}=\frac{\sqrt[3]{r_s^4\,l^2}}{2 r^4},
\end{equation}
and thus, despite the fractional powers of $\mathfrak{F}$ in the Lagrangian (\ref{denshayw}), the theory is well defined.

Note that, Hayward's line element (\ref{eq1}) is reduced to the Schwarzschild spacetime in the following manners: exact reduction for $ l = 0 $, and approximate reduction for $ r \gg l> 0$. Accordingly, the lapse function can be recast as
\begin{equation}
\label{eq3.1}
f(r)\simeq 1-\frac{r_s}{r}+\frac{r_s^2\,l^2}{r^4}-...,
\end{equation}
which is asymptotically flat. On the other hand, when $r \ll l$, the spacetime becomes similar to that of de Sitter, i.e.
\begin{equation}
\label{eq3.2}f(r)\simeq 1-\frac{r^2}{l^2}+\frac{r^5}{r_s\,l^4}-...,
\end{equation} 
and is flat and regular at $r=0$.

The possible horizons correspond to the zeros of the polynomial (\ref{eq3}). It is therefore necessary to study these roots in detail and obtain their expressions in terms of the parameters $\{r_s, l\}$.

\subsection{The nature of the roots}

\begin{figure}[h!]
	\begin{center}
		\includegraphics[width=84mm]{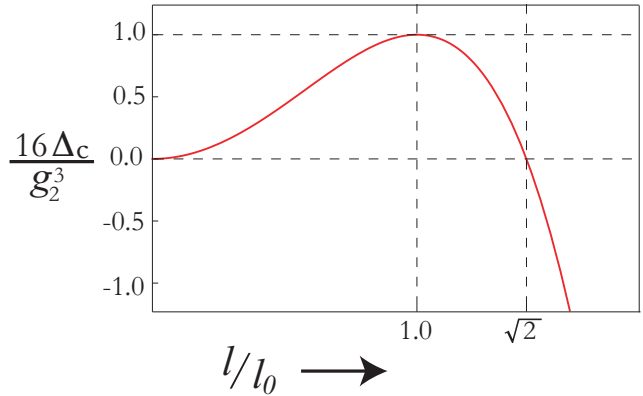}
	\end{center}
	\caption{The behavior of $\Delta_c$ (scaled by $g_2^3/16$) as a function of $l/l_0$, as given by Eq. (\ref{eq7}). The zeros are placed at $l_1/l_0=0$ and $l_2/l_0=\sqrt{2}\simeq 1.41$, and there is a maximum at $l/l_0=1$.}
	\label{fdisc}
\end{figure}

In order to have a qualitative approach to the nature of the aforementioned roots {by using the Cardano's method}, let us write the characteristic polynomial in its canonical form by defining $r = x+r_s/3$, so that
 \begin{equation}
 \label{eq4}P_3(x; r_s, l) \equiv x^3-(g_2/4) \,x-(g_3/4),
 \end{equation}where $g_2$ and $g_3$ are the Weierstra\ss \, invariants given by
 \begin{equation}
 \label{eq5} g_2=\frac{4 r_s^2}{3}>0,\,\, {\rm and}\,\,\,
 g_3=\frac{8\, r_s^3}{27}\,\left[1-\frac{l^2}{l_0^2}\right],
 \end{equation}respectively, and we have conveniently defined the quantity
 \begin{equation}
 \label{eq6}l_0\equiv \sqrt{\frac{2}{27}}\,r_s\approx 0.272\,r_s.
 \end{equation}
 Accordingly, the cubic  discriminant $\Delta_c=4(g_2/4)^3-27 (g_3/4)^2$, associated with the polynomial (\ref{eq4}) becomes 
 \begin{equation}
 \label{eq7}\Delta_c=\frac{g_2^3}{16}\,\left[1-\left(1-\frac{l^2}{l_0^2}\right)^2\right],
 \end{equation}
 vanishing at
 \begin{equation}
 \label{eq8} l_1=0,\quad {\rm and}\quad
 l_2=\sqrt{2}\,l_0\approx 0.385\,r_s.
 \end{equation}
 From the general theory of cubic polynomials (see for example Ref.~\cite{lang02}) we know that the sign of the discriminant determines the nature of the roots. Hence, the polynomial has
 \begin{enumerate}
 	\item three distinct real roots for $\Delta_c>0$ ($l_1<l<l_2$),
 	\item a multiple root for $\Delta_c=0$ ($l=l_1$ or $ l=l_2$),
 	\item one real root and a complex conjugate pair for $\label{eq10}\Delta_c<0$ ($l_2<l<\infty$).
 \end{enumerate}
Figure \ref{fdisc} shows the behaviour of the cubic discriminant (\ref{eq7}) as a function of the normalized Hayward's parameter $l/l_0$. The next section will be devoted to finding, analytically, the roots and discussing their physical interpretations.

\subsection{The roots}\label{roots22}

\begin{figure}[h!]
	\begin{center}
		\includegraphics[width=84mm]{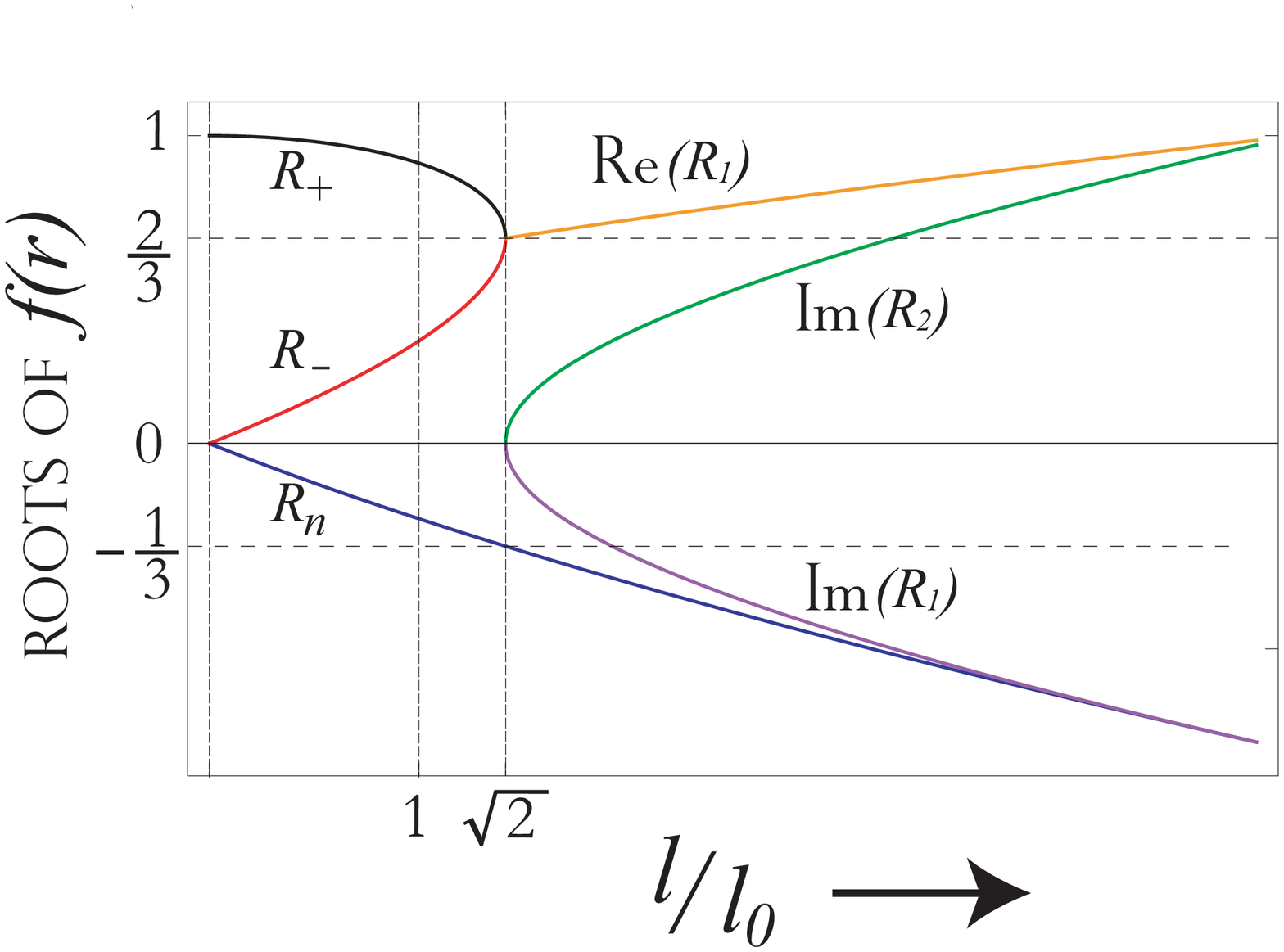}
	\end{center}
	\caption{Disposition of the roots of the Hayward's lapse function (per $r_s$)  as functions of Hayward's parameter $l$ (per $l_0$). For $0\leq l\leq l_0$ the roots are given by Eqs. (\ref{eq17}-\ref{eq19}), for $l_0\leq l\leq \sqrt{2}\,l_0$ by Eqs. (\ref{eq25}-\ref{eq27}), and for $\sqrt{2}\,l_0\leq l<\infty$ by Eqs. (\ref{eq31}-\ref{eq32}). 
	}
	\label{fraiz}
\end{figure}

\begin{figure}[h!]
	\begin{center}
		\includegraphics[width=84mm]{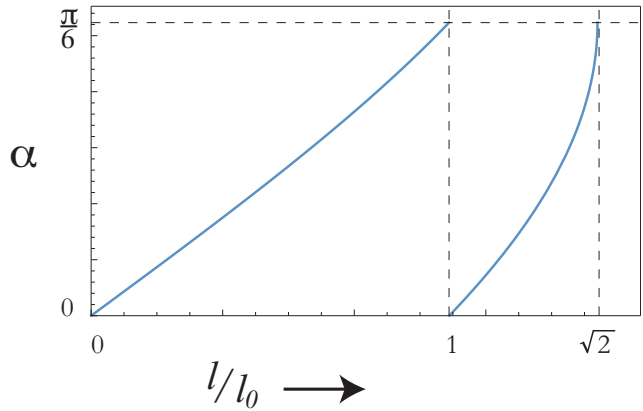}
	\end{center}
	\caption{The behavior of $\alpha$ as a function of $l/l_0$, based on Eq. (\ref{eq20}) for $l_1\leq l\leq l_0$, and on Eq. (\ref{eq28}) for $l_0\leq l\leq l_2$.}
	\label{falph}
\end{figure}

Since $r_s>0$ for all $M \in \mathbb{R}^+$, one can see directly from Eq. (\ref{eq5}) that the $g_3>0$ for $l<l_0$, $g_3<0$ for $l>l_0$, and $g_3=0$ for $l=l_0$. This means that the nature of the roots can be studied distinctly in terms of trigonometric and hyperbolic representations, depending on the value of the ratio $l/l_0$. Clearly, in the domain $l_1<l<l_0$, we have $g_3>0$ and the fundamental equation (\ref{eq4}) retains the standard form
\begin{equation}
\label{eq12} 4\,x^3-g_2\,x-g_3=0,
\end{equation}which can be compared with the trigonometric identity
\begin{equation}
\label{eq13} 4\,\cos^3\alpha-3\cos \alpha-\cos 3\alpha=0.
\end{equation}
Accordingly, applying some standard methods (see the appendices in Refs.~\cite{Cruz:2004ts,fkov} for details), and recalling that $r=x+r_s/3=x+\sqrt{g_2/12}$, we can write
\begin{eqnarray}
\label{eq17}R_+\equiv \frac{r_+}{r_s}&=&\frac{1+2\cos \alpha}{3},\\
\label{eq18}R_-\equiv \frac{r_-}{r_s}&=&\frac{1-\cos \alpha}{3}+ \frac{\sqrt{3}}{3}\sin\alpha,\\
\label{eq19}R_n\equiv \frac{r_n}{r_s}&=&\frac{1-\cos \alpha}{3}- \frac{\sqrt{3}}{3}\sin\alpha,
\end{eqnarray}
for $l_1\leq l\leq l_0$, where 
\begin{equation}
\label{eq20} \alpha=\frac{1}{3}\,{\rm arccos}\,\, \left(1-\frac{l^2}{l_0^2}\right).
\end{equation}
In the domain $l_0\leq l\leq l_2$, we get $g_3<0$ and thereby, the fundamental equation (\ref{eq4}) is recast as
\begin{equation}
\label{eq23} 4\,x^3-g_2\,x+|g_3|=0,
\end{equation}
and the employed identity will be
\begin{equation}
\label{eq24} 4\,\sin^3\alpha-3\sin \alpha+\sin 3\alpha=0.
\end{equation}
Following the same approach, it is straightforward to show that the three real roots are given by
\begin{eqnarray}
\label{eq25}
R_+\equiv \frac{r_+}{r_s}&=&\frac{1-\sin \alpha}{3}+ \frac{\sqrt{3}}{3}\cos\alpha,\\
\label{eq26}
R_-\equiv \frac{r_-}{r_s}&=&\frac{1+2\sin \alpha}{3},\\
\label{eq27}
R_n\equiv \frac{r_n}{r_s}&=&\frac{1-\sin \alpha}{3}- \frac{\sqrt{3}}{3}\cos\alpha,
\end{eqnarray}
where
\begin{equation}
\label{eq28}\alpha=\frac{1}{3}\arcsin\,\left(\frac{l^2}{l_0^2}-1\right).
\end{equation}
Finally, in the domain $l_2\leq l<\infty$, there is only one (negative) real root, $r_n$, and $r_1=r_2^*$. To find their explicit expressions, we first should note that in this domain the condition $l^2/l_0^2-1>1$ implies that the {\it angle} $\alpha$ in Eq.~(\ref{eq28}) is complex. Hence, using the identity
\begin{equation}
\label{eq29}\arcsin (x)=\frac{\pi}{2}-\imath\,\, {\rm arccosh} (x), \quad x\geq 1,
\end{equation}
in Eq. (\ref{eq28}), yields 
\begin{equation}
\label{eq30}\alpha=\frac{\pi}{6}-\frac{\imath}{3}{\rm arccosh}\,\left(\frac{l^2}{l_0^2}-1\right)\equiv \frac{\pi}{6}-\imath \,\beta.
\end{equation}
Therefore, applying Eq. (\ref{eq30}) in Eqs.~(\ref{eq25}-\ref{eq27}), and doing a brief manipulation, one gets
\begin{eqnarray}
\label{eq31}R_1\equiv \frac{r_1}{r_s}&=&\frac{1+\cosh \beta}{3}-\imath\, \frac{\sqrt{3}}{3}\sinh\beta=\frac{r_2^*}{r_s}\equiv R_2^*,\\
\label{eq32} R_n\equiv \frac{r_n}{r_s}&=&-\frac{2\cosh \beta-1}{3}.
\end{eqnarray}
The scheme in Fig. \ref{fraiz}, shows the behavior of the complete set of roots as functions of the ratio $l/l_0$. Furthermore, Fig. \ref{falph} indicates the change of the angle $\alpha$ as it changes in terms of the same ratio.
Note that, the solutions given by Eqs. (\ref{eq17}-\ref{eq18}) was already reported by Chiba \& Kimura in Ref.~\cite{chiba17} for the full domain $l_1\leq l\leq l_2$. Obviously, for $l_0\leq l\leq l_2$, their solutions are equivalent to those given in Eqs. (\ref{eq25}-\ref{eq27}), since $\cos \alpha>0$. However, $\cos 3\alpha <0$ and therefore, the condition given in Eq. (\ref{eq23}) is also satisfied.

Before going any further, let us review some general properties related to the obtained roots.

\paragraph{Proposition 1:}If $R_+(\alpha)$ is the function given by Eqs. (\ref{eq17}) or (\ref{eq25}) with $\alpha=\alpha (r_s,l)$ given by Eqs. (\ref{eq20}) or (\ref{eq28}), then in the domain $l_1\leq l\leq l_2$ we have
\begin{eqnarray}
\label{lemma1}\left(\frac{{\rm d} \,R_+}{{\rm d} \, \alpha}\right)&=& f_*(\alpha),\\
\label{lemma}r_s\left(\frac{\partial \,\alpha}{\partial \, r_s}\right)_{l}&=&-l\left(\frac{\partial \, \alpha}{\partial \, l}\right)_{r_s}=h_*(\alpha),\\
\label{lemma3}r_s\left(\frac{\partial \,R_+}{\partial \, r_s}\right)_{l}&=&-l\left(\frac{\partial \, R_+}{\partial \, l}\right)_{r_s}=g_*(\alpha)=f_*\, h_*,
\end{eqnarray}
where the functions $f_*(\alpha), g_*(\alpha)$ and $h_*(\alpha)$ are given by
\begin{eqnarray}
\label{f1}f_*(\alpha)&=&-\frac{2}{3}\sin \alpha,\\
\label{h1}h_*(\alpha)&=&-\frac{2}{3}(\csc 3\alpha-\cot 3\alpha),\\
\label{g1}g_*(\alpha)&=&\frac{4}{9}\,\sin\alpha \left(\csc 3\alpha-\cot 3\alpha \right),
\end{eqnarray} 
for $l_1\leq l\leq l_0$, and
\begin{eqnarray}
\label{f2}f_*(\alpha)&=&-\frac{1}{3}(\cos\alpha+\sqrt{3}\sin \alpha),\\
\label{h2}h_*(\alpha)&=&-\frac{2}{3}(\sec 3\alpha+\tan 3\alpha),\\
\label{g2}g_*(\alpha)&=&\frac{2}{9}\, (\cos\alpha+\sqrt{3}\sin\alpha)\left(\sec 3\alpha+\tan 3\alpha
\right),
\end{eqnarray}
for $l_0\leq l \leq l_2$.

\section{The ($r_s, l$) thermodynamics}\label{termasp}

In this section, we discuss some thermodynamic consequences of the obtained relations, and in particular, their relevance to the study of the black hole's entropy, using the Bekenstein-Hawking approach. This way, once the entropy of the system has been calculated, we can obtain other thermodynamic quantities of interest. In what follows, we treat the Hayward black hole as a thermodynamic system, by considering the pair ($r_s, l$) as the thermodynamic variables, {which are extensive and homogeneous of degree one (see Sect. \ref{ap1})}.

\subsection{The basis}
\begin{figure}[h!]
	\begin{center}
		\includegraphics[width=84mm]{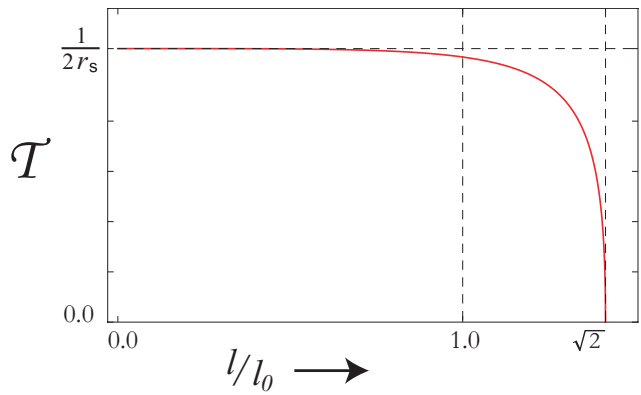}
	\end{center}
	\caption{Temperature parameter $\mathcal{T}$ as a function of the dimensionless ratio $l/l_0$.}
	\label{ftemp}
\end{figure}

\begin{figure}[h!]
	\begin{center}
		\includegraphics[width=84mm]{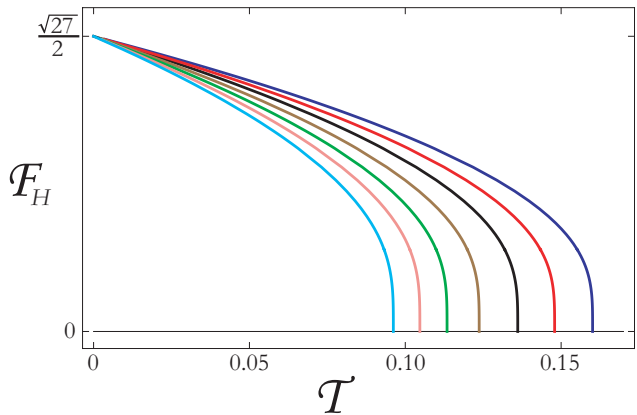}
	\end{center}
	\caption{The Hayward's force as a function of the temperature parameter $\mathcal{T}$. {All curves converge to  $\mathcal{F}_H=\sqrt{27}/2$ at the thermodynamic limit $\mathcal{T}\rightarrow 0$, which corresponds to the extreme black hole. This is while for $\mathcal{F}_H\rightarrow0$, each curve tends to its isotherm state $\mathcal{T}\rightarrow\mathcal{T}_s$.}}
	\label{ft}
\end{figure}

\begin{figure}[h!]
	\begin{center}.
		\includegraphics[width=84mm]{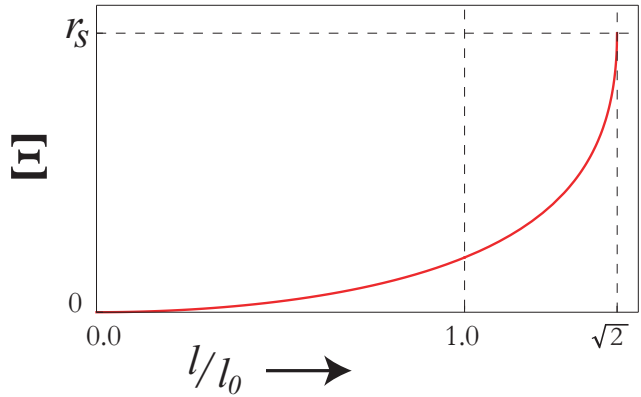}
	\end{center}
	\caption{ The Hayward's potential $\Xi$ given by Eq. (\ref{pothaw}), as a function of $l/l_0$. 
	}
	\label{fhaypot}
\end{figure}

{We consider the Bekenstein-Hawking approach to study the thermodynamics of Hayward black hole \cite{Bekenstein:1972,Bekenstein:1973,Bekenstein:1974,Bekenstein:1975,Bardeen:1973gs,Hawking:1974sw}. It is necessary to keep in mind that through this procedure, the construction of the thermodynamics is implicitly established in the Carathéodory framework, in which, the existence of a function, termed as {\it metrical entropy}, is ensured along with the {\it absolute temperature} (see Sect. \ref{ap1}). This will be delved deeper in a future work \cite{fmv20}. Accordingly, one could connect the Carathéodory and Gibbs frameworks through homogeneity, as established by Belgiorno \cite{Belgiorno2002,Belgiorno:2002iv}. We therefore adopt this approach based on the following postulate: {\it the event horizon area of a black hole cannot decrease; it increases in most of the transformations of the black hole} \cite{Misner:1974qy}. Thus, letting $A=4\pi r_+^2$ to be the area of the event horizon, the entropy is then given by}

\begin{equation}
\label{entropy} 
S=\frac{k_B}{4}\frac{4 \pi r_+^2}{\ell_p^2},
\end{equation}
where $k_B$ is the Boltzmann constant and $\ell_p$ is the Planck length. It is therefore useful to define the entropy function as
\begin{equation}
\label{entrfcn}\mathcal{S}(r_s, l)\equiv \frac{4\,\ell_p^2\,S}{4 \pi k_B}=r_s^2\,R_+^2(r_s, l).
\end{equation}
which is used hereinafter. Note that, $r_s$ represents the energy-mass parameter, {which here is considered to be the} internal energy $U$, and $\mathcal{S}$ has units of area with the SI conversion factor $1.32\times10^{16}\,$[J/(fm$^2$\,K)]. Now, since $\mathcal{S}\equiv\mathcal{S}(r_s, l)$, it follows that
\begin{equation}
\label{difentr}{\rm d}\mathcal{S}=\left(\frac{\partial \mathcal{S}}{\partial r_s}\right)_l\,{\rm d} r_s + \left(\frac{\partial \mathcal{S}}{\partial l}\right)_{r_s}\,{\rm d} l,
\end{equation}
which allows to define the temperature parameter $\mathcal{T}$ as
\begin{equation}
\label{temp}\frac{1}{\mathcal{T}}=\left(\frac{\partial \mathcal{S}}{\partial r_s}\right)_l,
\end{equation}or, using Eqs. (\ref{lemma1}-\ref{lemma3}),
\begin{equation}
\label{teml2}\mathcal{T}=\frac{\mathcal{T}_s}{R_+(\alpha)\,[R_+(\alpha)+g_*(\alpha)]},
\end{equation}
where
\begin{equation}
\label{templ3}\mathcal{T}_s=\frac{1}{2r_s},
\end{equation}
is the temperature corresponding to the Schwarzschild black hole. As we can see from Fig. \ref{ftemp}, for small values of $l$, the temperature remains near $\mathcal{T}_s$. In this case, we can expand $\mathcal{T}$ as
\begin{equation}
\label{tlch}\mathcal{T}\simeq \mathcal{T}_s\,\left(1-\frac{3 l^4}{r_s^4}- \frac{20 l^6}{r_s^6}+...\right), \quad l\ll r_s.
\end{equation}
On the other hand, for $l\rightarrow l_2$, we have that the temperature rapidly drops to zero as $g_*(\alpha)$ approaches goes to infinity. Accordingly, 
\begin{equation}
\label{tcero}\mathcal{T}\simeq \frac{\mathcal{T}_s}{R_+(\alpha)\,g_*(\alpha)}\left(1-\frac{R_+(\alpha)}{g_*(\alpha)}+...\right),\qquad \,\,l\lesssim l_2,
\end{equation}
which corresponds to the low temperature limit.
Additionally, we treat the Hayward's parameter $l$  as a generalized displacement associated with a generalized mechanical force, $\mathcal{F}_H$, defined as
\begin{equation}
\label{hf}\mathcal{F}_H\equiv -\mathcal{T}\,\left(\frac{\partial \mathcal{S}}{\partial l}\right)_{r_s}=\frac{\sqrt{27}}{2}\,\frac{g_*(\alpha)}{R_+(\alpha)+g_*(\alpha)}\,\frac{l_2}{l}. 
\end{equation}{Since $l$ is related to the magnetic charge $q_m$ (cf. with Eq. (\ref{magchaHa})), the Hayward's force is related to the magnetic potential, which is the conjugate canonical variation of $q_m$.}
{In Fig. \ref{ft}, we have plotted the behavior of $\mathcal{F}_H$ as a function of the temperature $\mathcal{T}$. Each particular curve corresponds to a fixed value of $r_s$, and the Hayward's parameter varies in the interval $l_1\leq l\leq l_2$. Note that for the extreme black hole, all of these curves converge to the value $\mathcal{F}_H=\sqrt{27}/2$. On the other hand, when the Hayward's force vanishes, the curves tend to their corresponding temperature $\mathcal{T}_s$,}
\begin{eqnarray}
\label{asintft2}\mathcal{T}\rightarrow 0&&\Rightarrow \mathcal{F}_H\rightarrow \sqrt{27}/2,\\
\label{asintft}\mathcal{T}\rightarrow \mathcal{T}_s&&\Rightarrow \mathcal{F}_H\rightarrow 0.
\end{eqnarray}
{In turns, the introduction of the conjugated canonical pair ($l, \mathcal{F}_H$) allows us to define the {\it Hayward's free energy}, $\Xi$, through the following relation:} 
\begin{equation}
\label{pothaw}\Xi\equiv \mathcal{F}_H l=\frac{g_*(\alpha)}{R_+(\alpha)+g_*(\alpha)}\,r_s.
\end{equation} 
{As can be seen from Fig. \ref{fhaypot}, for small values of $l$, this thermodynamic potential behaves as}
\begin{equation}
\label{pothaylim}\Xi \simeq \frac{2 l^2}{r_s}\,\left(1+\frac{3 l^2}{r_s^2}+\frac{12 l^4}{r_s^4}+...\right),
\end{equation} 
for $l\ll r_s$, and as
\begin{equation}
\label{pothaylim2}\Xi\simeq r_s\,\left(1-\frac{R_+(\alpha)}{g_*(\alpha)}+...\right),
\end{equation}
for $l\rightarrow l_2$.
As inferred from Eqs. (\ref{eq17}) and (\ref{g1}), in the limit  $l/l_0\rightarrow0$, we have $R_+\rightarrow1$ and $g_*\rightarrow0$, and hence, $\Xi\rightarrow0$. On the other hand, for $l/l_0\rightarrow\sqrt{2}$, we have that $R_+\rightarrow 2/3$ and $g_*\rightarrow\infty$, so $\Xi\rightarrow r_s$. Thus, the free energy is minimum for the Schwarzschild case ($l=l_1$) and maximum for the extreme Hayward black hole ($l=l_2$).
{Since $l$ is related to the magnetic charge and $\mathcal{F}_{H}$ to the magnetic potential, the Hayward's potential $\Xi$ is related to the magnetic energy of the source.}

\begin{figure}[h!]
	\begin{center}.
		\includegraphics[width=100mm]{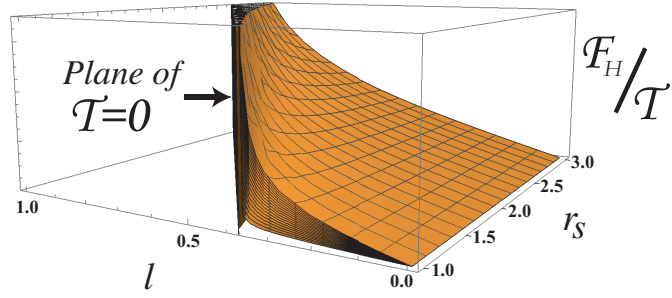}
	\end{center}
	\caption{ {The EoS for the Hayward black hole, as given by Eq. (\ref{eos}). Each foliation is a surface of $\mathcal{T}=$ const. The upper and lower surfaces correspond respectively to $\mathcal{T}=10$ and $\mathcal{T}=1$ (in arbitrary units).}
	}
	\label{figeos}
\end{figure}

{Now, combining the first and second laws of thermodynamics, we can write
\cite{reichl2009modern}}
\begin{equation}
\label{primseg}\mathcal{T}\,{\rm d}\mathcal{S}\geq {\rm d}r_s-\mathcal{F}_H\,{\rm d}l,
\end{equation}
{for which, the equality holds for reversible processes. 
It can be easily noticed from Eq. (\ref{entrfcn}), that the entropy is a homogeneous function of the second degree in $r_s$ and  $l$. Accordingly, the Euler's theorem requires}
\begin{equation}
\label{fundeq} 2\, \mathcal{T} \mathcal{S} =r_s-\mathcal{F}_H\,l=r_s-\Xi.
\end{equation}
This is the {\it fundamental equation} or the {\it Gibbs-Duhem relation} for the Hayward black hole. {Also, the {\it equation of state} (EoS) relating the thermal state variable, $\mathcal{T}$, to the mechanical variables of the system, can be obtained by combining Eqs. (\ref{teml2}) and (\ref{hf}), as
\begin{equation}
\label{eos}\frac{\mathcal{F}_H}{\mathcal{T}}=2\frac{ r_s^2\, g_*(r_s, l) \, R_+(r_s, l)}{l}.
\end{equation} {In Fig. \ref{figeos} we have plotted the EoS in the following way: each of the foliations corresponds to a surface of the EoS on which $\mathcal{T}= $ const., and the extensive variables ($r_s, l$) do vary. This way, an hypervolume is generated by adding more foliations, through which, the system can move. It is important to note that, each curve tends asymptotically to the plane of $\mathcal{T}=0$. This means that their corresponding states are disconnected from those of $\mathcal{T}\neq0$. This is in agreement with the third law's requirement \cite{Belgiorno:2002pm,Belgiorno_2003a,Belgiorno_2003b}.}
} 

{Note that, the mass-energy $ r_s $ can be treated as free energy, in the sense that the work can be stored in the form of potential energy and be recovered later. 
In fact, the total mass-energy differential can be rewritten from Eq. (\ref{primseg}) as}
\begin{equation}
\label{primseg2} {\rm d}r_s\leq \mathcal{T}\,{\rm d}\mathcal{S}+\mathcal{F}_H\,{\rm d}l,
\end{equation}
{where the inequality holds for spontaneous changes. Hence, for reversible changes, we can infer from Eq. (\ref{primseg2}) that}
\begin{equation}
\label{du1}\mathcal{T}=\left(\frac{\partial r_s}{\partial \mathcal{S}}\right)_l,
\end{equation}
\begin{equation}
\label{du2}\mathcal{F}_H=\left(\frac{\partial r_s}{\partial l}\right)_{\mathcal{S}},
\end{equation}
{from which, one of the Maxwell relations is obtained as}
\begin{equation}
\label{max1}\left(\frac{\partial \mathcal{T}}{\partial l}\right)_{\mathcal{S}}=\left(\frac{\partial \mathcal{F}_H}{\partial \mathcal{S}}\right)_l.
\end{equation}
{If we consider the black hole as a closed and isolated system, we can write}
\begin{equation}
\label{deltu}\Delta r_s=\Delta Q-\Delta W,
\end{equation}
for changes in the black hole's mass-energy, {where $\Delta Q$ is the heat flux across the surface of the black hole, and $\Delta W$ is the work that can be divided into two parts: the first part, $\int \mathcal{F}_H\mathrm{d}l$, changes the parameter $l$, and the second part, $\Delta W_{free}$, is the extra work performed by the system.}
\begin{equation}
\label{trabrs}\Delta W=\int \mathcal{F}_H{\rm d}l+\Delta W_{free}.
\end{equation}
{Then, in a reversible process, with $\mathcal{S}$ and $l$ as constants, we have that}
\begin{equation}
\label{delrs1}(\Delta r_s)_{\mathcal{S}, l}=-\Delta W_{free},
\end{equation}
{and in this way, the work can be stored in the form of mass-energy and then completely restored. If in this process there is no work done by or on the system, we have}
\begin{equation}
\label{delrs2}(\Delta r_s)_{\mathcal{S}, l}=0,
\end{equation}
{so that the mass-energy remains unchanged. Since a system in equilibrium cannot change its state spontaneously, the equilibrium state with fixed $ \mathcal{S}$ and $l$ has the minimum mass-energy $r_s$.}

{We can now express the corresponding enthalpy as}
\begin{equation}
\label{ental} \mathcal{H}\equiv r_s-\mathcal{F}_H\,l=2\, \mathcal{T} \mathcal{S},
\end{equation}
which is obtained by adding an additional energy, that accounts for a mechanical coupling. {Using Eqs. (\ref{lemma1}-\ref{lemma3}), we get}
\begin{equation}
\label{ental2} \mathcal{H}=r_s\,\frac{R_+(\alpha)}{R_+(\alpha)+g_*(\alpha)}.
\end{equation}
{Accordingly, taking the differential of Eq. (\ref{ental}) and then combining it with Eq. (\ref{primseg}), we obtain}
\begin{equation}
\label{difental}{\rm d}\mathcal{H}\leq \mathcal{T} {\rm d}\mathcal{S}-l {\rm d}\mathcal{F}_H,
\end{equation}
{with which, we can write}
\begin{equation}
\label{dh1}\mathcal{T}=\left(\frac{\partial \mathcal{H}}{\partial \mathcal{S}}\right)_{\mathcal{F}_H},
\end{equation}
\begin{equation}
\label{dh2}l=-\left(\frac{\partial \mathcal{H}}{\partial \mathcal{F}_H}\right)_{\mathcal{S}},
\end{equation}
{for reversible processes, and hence, the new Maxwell relation}
\begin{equation}
\label{max2}\left(\frac{\partial \mathcal{T}}{\partial \mathcal{F}_H}\right)_{\mathcal{S}}=-\left(\frac{\partial l}{\partial \mathcal{S}}\right)_{\mathcal{F}_H},
\end{equation}
\begin{figure}[h!]
	\begin{center}
		\includegraphics[width=75mm]{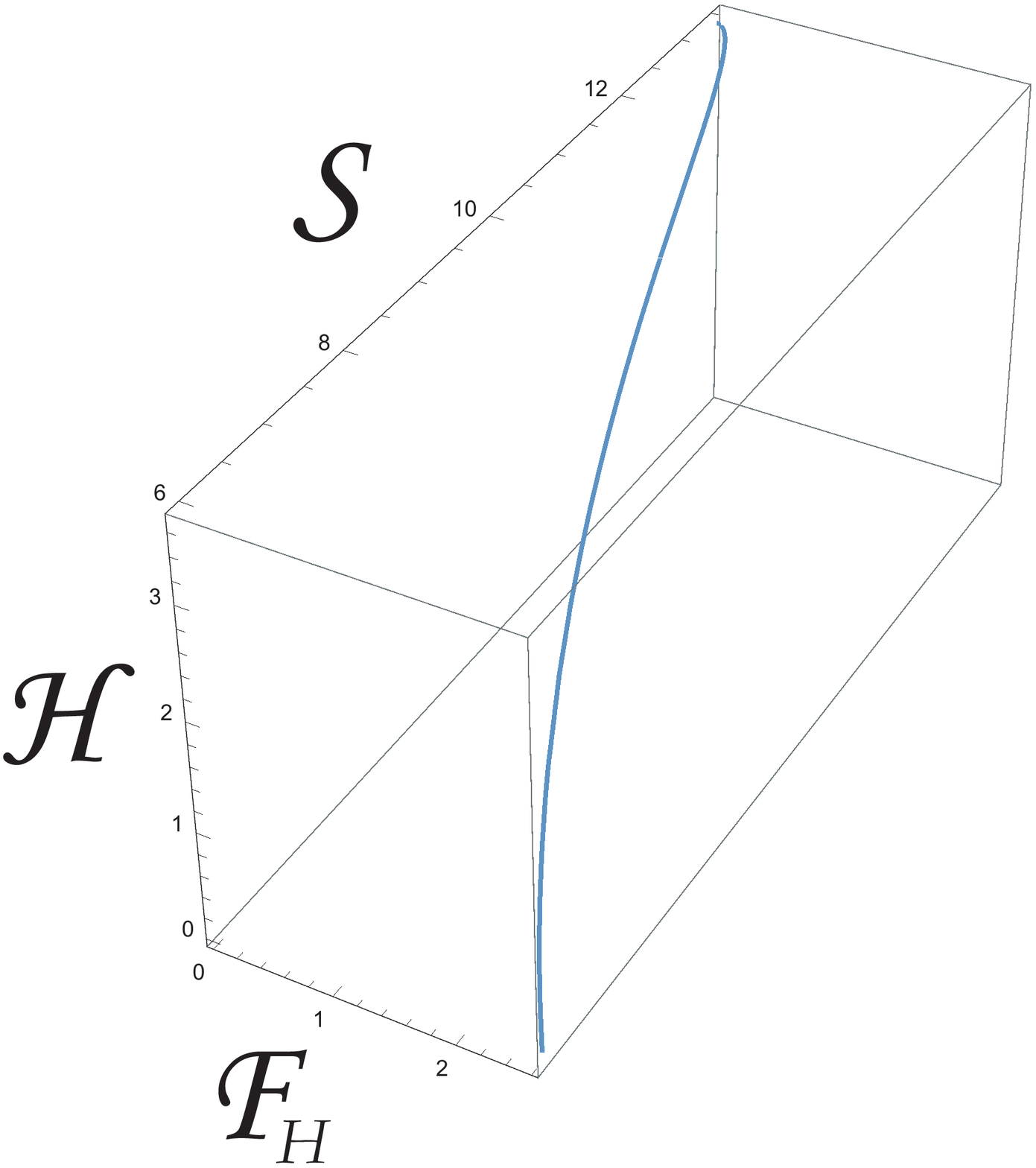}
		\includegraphics[width=83mm]{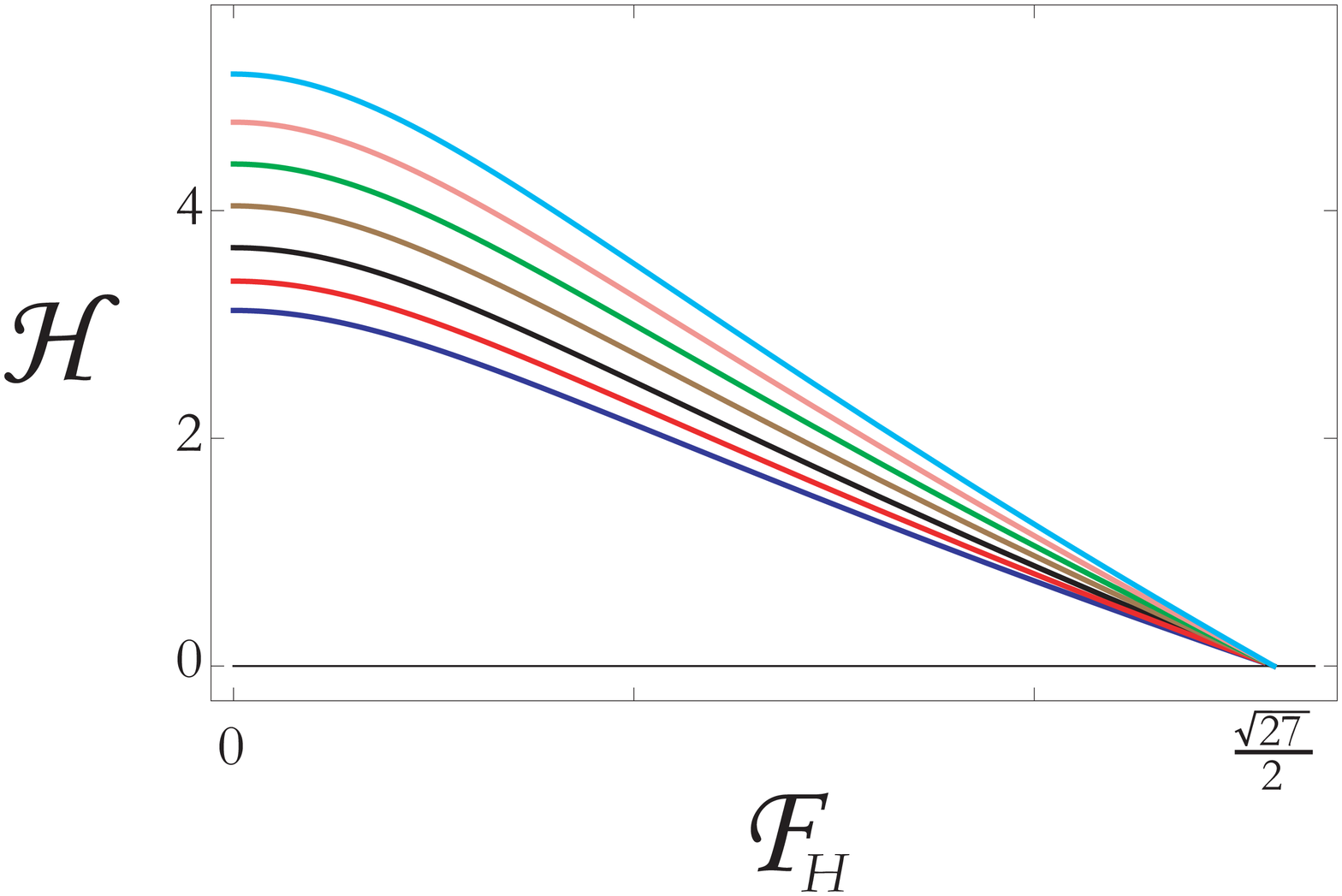}
	\end{center}
	\caption{Top panel: A typical curve on the $(\mathcal{F}_H, \mathcal{S}, \mathcal{H})$ space; Bottom panel: The enthalpy $\mathcal{H}$ as a function of the Hayward's force $\mathcal{F}_H$.
	}
	\label{fent}
\end{figure}is inferred. The top panel of Fig. \ref{fent} shows the behavior of enthalpy as a function of its natural variables ($\mathcal{S}, \mathcal{F}_H$), while the bottom panel of same figure, shows the projection on the $\mathcal{H}-\mathcal{F}_H$ plane for several of these curves. It is not difficult to prove that at the limit of the Schwarzschild black hole, the enthalpy takes the known value $r_s$ for zero cosmological constant \cite{Kastor:2009wy}, while for the extreme Hayward black hole, the enthalpy vanishes. To summarize,
\begin{eqnarray}
\label{entschw}\mathcal{F}_H\rightarrow 0\quad &\Rightarrow& \mathcal{H}\rightarrow r_s,\\
\label{entextr}\mathcal{F}_H\rightarrow \frac{\sqrt{27}}{2}\quad &\Rightarrow& \mathcal{H}\rightarrow 0.
\end{eqnarray}
In addition, using arguments similar to those used for energy-mass, it is possible to show that the equilibrium state of the Hayward black hole, with fixed $\mathcal{S}$ and $\mathcal{F}_H$, is a state of minimum enthalpy.

\begin{figure}[h!]
	\begin{center}
		\includegraphics[width=84mm]{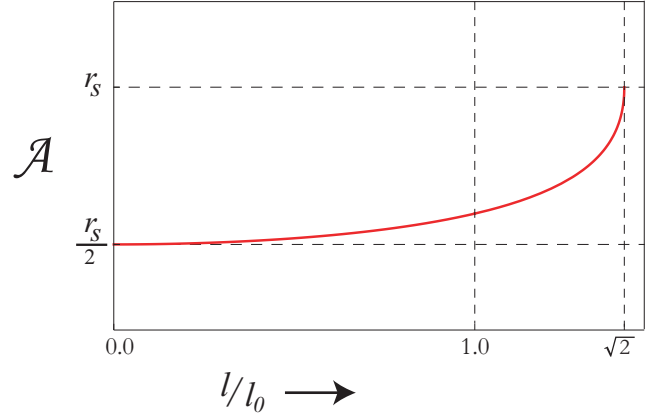}
	\end{center}
	\caption{The variation of the Helmholtz free energy parameter $\mathcal{A}$ as a function of $l/l_0$. This energy has a minimum $r_s/2$ at $l=0$, and a maximum $r_s$ at $l=\sqrt{2}l_0$.}
	\label{fhelm}
\end{figure}

Finally, we obtain the Helmholtz free energy parameter from the energy-mass $r_s$, by adding a term due to the thermal coupling. We have
\begin{equation}
\label{helm1}\mathcal{A}\equiv r_s-\mathcal{T}\,\mathcal{S}=\mathcal{T}\,\mathcal{S}+\Xi,
\end{equation} 
which using Eqs. (\ref{lemma1}-\ref{lemma3}),  becomes
\begin{equation}
\label{helml2}
\mathcal{A}=\frac{r_s}{2}\,\left[\frac{R_+(\alpha)+2g_*(\alpha)}{R_+(\alpha)+g_*(\alpha)}\right].
\end{equation}
{As shown in Fig. \ref{fhelm}, this quantity has a minimum $\mathcal{A}_1=r_s/2$ at $l=l_1=0$, and a maximum $\mathcal{A}_2=r_s=2\mathcal{A}_1$ at $l=l_2=\sqrt{2}l_0$.}
Taking the differential of Eq. (\ref{helm1}) and then using  Eq. (\ref{difentr}), we find
\begin{equation}
\label{difhelm}{\rm d}\mathcal{A}=-\mathcal{S}\,{\rm d}\mathcal{T}+\mathcal{F}_H\,{\rm d}l,
\end{equation}
and therefore,
\begin{eqnarray}
\label{as1}\mathcal{S}&=&-\left(\frac{\partial \mathcal{A}}{\partial \mathcal{T}}\right)_l,\\
\label{af1}\mathcal{F}_H&=&\left(\frac{\partial \mathcal{A}}{\partial l}\right)_{\mathcal{T}}.
\end{eqnarray}
{As before, we can obtain the corresponding Maxwell relation as}
\begin{equation}
\label{max3}\left(\frac{\partial \mathcal{F}_H}{\partial \mathcal{T}}\right)_{l}=-\left(\frac{\partial \mathcal{S}}{\partial l}\right)_{\mathcal{T}},
\end{equation}
{which supports the fact that the equilibrium state of the Hayward black hole, with fixed $\mathcal{T}$ and $l$, has the minimum Helmholtz free  energy. The corresponding Gibbs free energy will be presented in Subsect. \ref{tpt}.}

\subsection{The $\mathcal{T}-\mathcal{S}$ diagram}\label{tsd}

\begin{figure}[h!]
	\begin{center}
		\includegraphics[width=84mm]{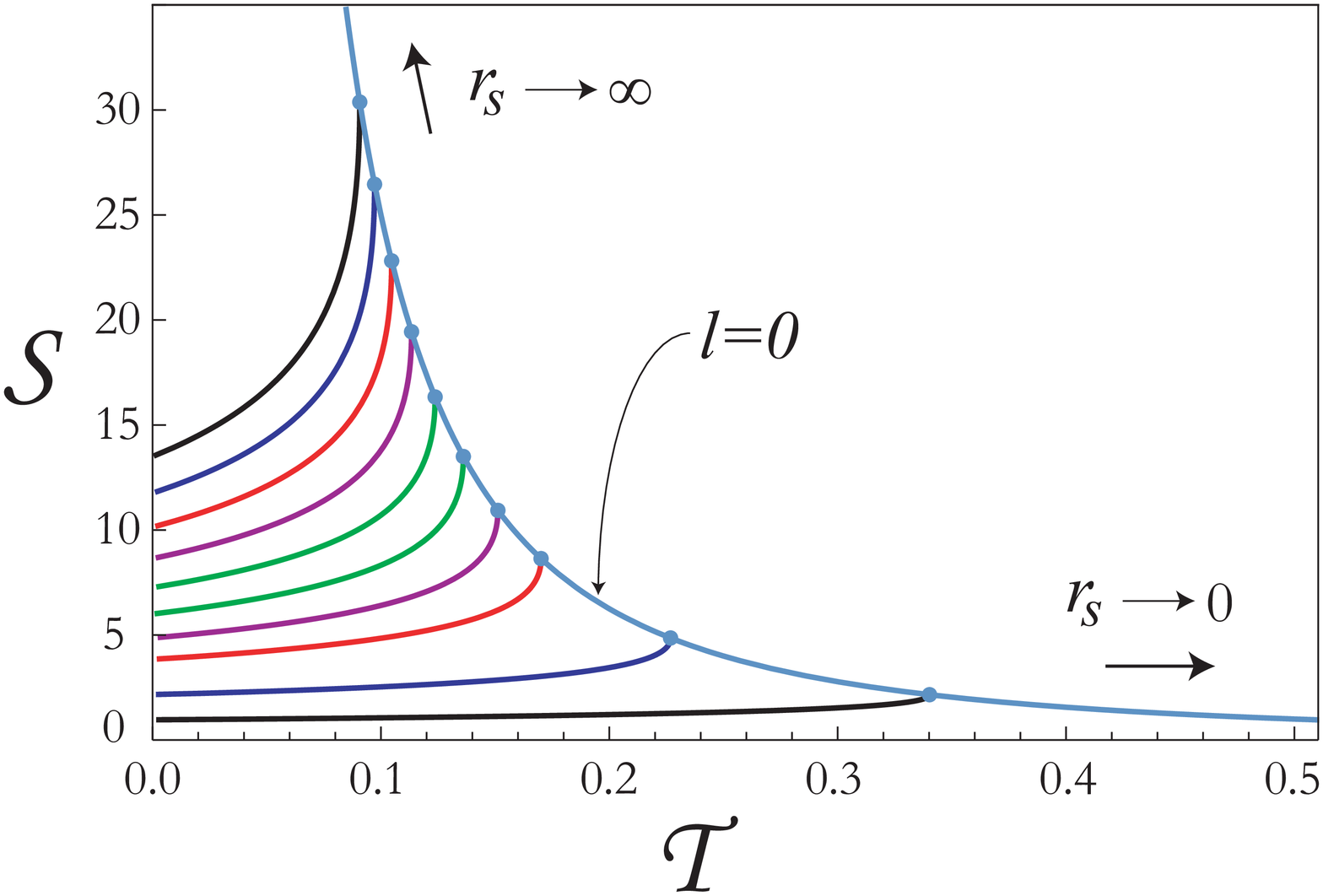}
	\end{center}
	\caption{The $\mathcal{S}-\mathcal{T}$ diagram according to Eqs. (\ref{entropy}) and (\ref{teml2}). The values of $l_0$ correspond to the chosen $r_s$, given as $l_0=\sqrt{2/27}\,r_s=0.4, 0.6, 0.8, 0.9, 1, 1.1, 1.2, 1.3, 1.4, 1.5$. All curves have the minimum value for $l=l_2$ (at $\mathcal{T}=0$) and raise to their maximum value for $l=0$ (the Schwarzschild black hole), whose state is characterized by the curve $\mathcal{S}_s=(4\mathcal{T}_s^2)^{-1}=r_s
^2$.}
	\label{fts}
\end{figure}

{Based on what we have discussed so far, the explicit expressions for $\mathcal{S}$ and $\mathcal{T}$ in terms of ($r_s, l$), make it possible to study the $\mathcal{S}-\mathcal{T}$ diagram, parametrically. This has been shown in Fig. \ref{fts}. Note that, the $ l = 0 $ curve (Schwarzschild black hole), corresponds to}
\begin{equation}
\label{ts0}\mathcal{S}_s\equiv\frac{1}{4\mathcal{T}_s^2}=r_s^2.
\end{equation}
{It is important to note that $\Delta \mathcal{S} >0$ implies that $\Delta l <0$ for all possible values of $r_s$. This means that the changes in $ l $, lead the Hayward black hole towards that of Schwarzschild.}

{One could also recall the conformity with the third law of thermodynamics, which rejects the possibility of reaching $\mathcal{T} = 0$ in a finite number of steps. In fact, if $r_s$ is fixed, $l$ grows necessarily and the area decreases \cite{Bardeen:1973gs}. On the other hand, it is important to highlight that}
\begin{equation}
    \label{limts}\mathcal{S}_0 \equiv \lim_{\mathcal{T}\rightarrow0}\mathcal{S}=\frac{4}{9}\,r_s^2=\frac{4}{9}\,\mathcal{S}_s,
\end{equation}
{which agrees with the Nerst-Planck postulation of the third law ($\mathcal{S}\rightarrow0$ as $\mathcal{T}\rightarrow0$), only when $r_s\rightarrow 0$ (see Refs.~\cite{Belgiorno:2002pm,Belgiorno_2003a,Belgiorno_2003b} for a good discussion on the third law).}

\subsection{The phase transition}\label{tpt}

\begin{figure}[h!]
	\begin{center}
		\includegraphics[width=80mm]{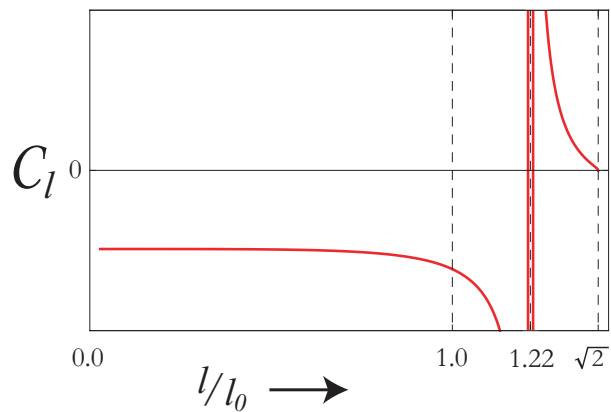}
	\end{center}
	\caption{The heat capacity parameter $\mathcal{C}_l$ at constant Hayward's parameter $l$ as a function of $l/l_0$. For the Schwarzschild case, $l=0$ and  $\mathcal{C}_l$ reduces to $-2 \,r_s^2$. The broken line  at $l=l_{pt}\simeq1.22\,l_0$ indicates the position of the phase transition for which, $\mathcal{C}_l$ experiences an infinite discontinuity. After this line, the heat capacity is positive, falling to zero at the extreme Hayward's value $l=l_2$, which corresponds to the thermodynamic limit.}
	\label{fcl}
\end{figure}

An important aspect of the thermodynamic systems emerges from varying the specific heats. In the particular case studied here, we consider the thermal capacity parameter as
\begin{equation}
\label{thercap}\mathcal{C}_l=\mathcal{T}\,\left(\frac{\partial \mathcal{S}}{\partial \mathcal{T}}\right)_l=\left(\frac{\partial \,r_s}{\partial \mathcal{T}}\right)_l=\left(\frac{\partial \mathcal{T}}{\partial \,r_s}\right)_l^{-1},
\end{equation}
by fixing $l$.
Basically, we suppose that the black hole is held at the temperature $\mathcal{T}$ and in thermal equilibrium with its environment. Then, if the environment is increased in temperature, the black hole absorbs the resultant energy. This way, the environment remains isotropic, such that there will be no increase in angular momentum and no changes in the parameter $l$. Therefore, using Eq. (\ref{teml2}) in Eq. (\ref{thercap}), we can write
\begin{equation}
\label{celc}\mathcal{C}_l=-\frac{r_s}{\mathcal{T}\,[1+Y(\alpha)]},
\end{equation}
where the function $Y(\alpha)$ is given by
\begin{equation}
\label{yfunc}Y(\alpha)=\frac{g_*(\alpha)}{R_+(\alpha)}+\frac{g_*(\alpha)}{R_+(\alpha)+g_*(\alpha)} \left[\frac{\partial \log (r_s\,g_*)}{\partial \log r_s}\right]_{l}.
\end{equation}
The general behavior of $\mathcal{C}_l$ as a function of the ratio $l/l_0$, at fixed $r_s$, is shown in Fig. \ref{fcl}. Note that, for the Schwarzschild black hole,  $Y(\alpha)=0$, and therefore
\begin{equation}
\label{clshw}\mathcal{C}_l=-\frac{r_s}{\mathcal{T}}=-\frac{1}{2}\frac{1}{\mathcal{T}^2}=-2r_s^2.
\end{equation}
This means that the Schwarzschild black hole gets hotter as it radiates energy \cite{davies77}. Another interesting feature is the infinite discontinuity at $l_{pt}$, after which, $\mathcal{C}_l$ changes from negative values to positive ones.
The critical point where the infinite discontinuity occurs, is the positive root of the transcendental equation
\begin{equation}
\label{trasceq}1+Y(\alpha(r_s, l_{pt}))=0,
\end{equation}
which results in $l_{pt}\simeq 1.22\,l_0\simeq 0.333\,r_s$, and therefore, $l_0<l_{pt}<l_2$. Note also that, the set $\{\mathcal{S}, \mathcal{T}, \mathcal{F}_H\}$ remains finite and continuous at the critical point $l_{pt}$. 

The corresponding Gibbs free energy parameter $\mathcal{G}$ is obtained by adding thermal and mechanical coupling terms to the energy-mass $r_s$, yielding
\begin{equation}
\label{gib1}\mathcal{G}=r_s-\mathcal{T} \mathcal{S}-\mathcal{F}_H\,l=\mathcal{A}-\Xi,
\end{equation} 
from which, we obtain
\begin{equation}
\label{difgib}{\rm d}\mathcal{G}= -\mathcal{S}\,{\rm d}\mathcal{T}-l\,{\rm d}\mathcal{F}_H,
\end{equation}
and therefore
\begin{equation}
\label{gradgib}\mathcal{S}=-\left(\frac{\partial \mathcal{G}}{\partial \mathcal{T}}\right)_{\mathcal{F}_H},\quad l=-\left(\frac{\partial \mathcal{G}}{\partial \mathcal{F}_H}\right)_{\mathcal{T}}.
\end{equation}
Clearly, the Gibbs free energy parameter (\ref{gib1}) and its first derivatives (\ref{gradgib}), are continuous at $l_{pt}$, but presents a discontinuity in the second derivatives (heat capacity). We are therefore encountering with a phase transition of the second order.

It is of worth to point out that, the behavior of the heat capacity in plotted in Fig. \ref{fcl}, is similar to that for the Kerr-Newman black hole, considering $C_{J, Q}$, $C_{\Omega, Q}$ and $C_{J, \Phi}$  to be functions of $Q/M$ or $J/M^2$ \cite{davies77,ruppeiner08}. {Furthermore, as in case of the Schwarzschild black hole, by radiating energy, the Hayward black hole gets hotter for $0\leq l<l_{pt}$ and gets colder for $l_{pt}<l\leq \sqrt{2} l_0$.}

{\subsection{A short note on the homogeneity}\label{ap1}
	Here,  based on Refs. \cite{Belgiorno:2002iv,Belgiorno:2002iw}, we present the first approach to the study of the homogeneity. A deeper study of this topic, along with other relevant concepts, will be dealt with in the second part of this paper \cite{fmv20}. 
	To begin, recall that the function $\alpha$ defined by Eqs. (\ref{eq20}) and (\ref{eq28}), is homogeneous of degree zero in terms of the independent variables $\{r_s, l\}$, i.e., $\alpha(\lambda r_s, \lambda l)=\alpha(r_s, l)$. The same holds for $R_+$, given by Eqs. (\ref{eq17}) and (\ref{eq25}), and the set of functions defined by Eqs. (\ref{lemma})-(\ref{g2}). Inspection of Eq. (\ref{entrfcn}) shows that by rescaling $(r_s, l)\mapsto (\lambda r_s, \lambda l)$, one gets $\mathcal{S} \mapsto \lambda^2 \mathcal{S}$, i.e., the entropy function is homogeneous of degree two. In the same way, one finds from Eq. (\ref{teml2}) that $\mathcal{T}$ is homogeneous of degree $-1$ and therefore, is not intensive. It is seen that, by doubling the Schwarzschild radius and the Hayward's parameter, the temperature becomes one-half of the temperature of the initial state of the black hole. On the other hand, the Hayward force (\ref{hf}) is homogeneous of degree zero, and hence, it is an intensive variable given in terms of the extensive variables as $\mathcal{F}_H(r_s, l)$. This statement indicates that $(l,\mathcal{F}_{H})$ is a thermodynamic canonical pair.
	In order to give more clarifications, let us consider the Euler vector field (Liouville operator) as
	\begin{equation}
	\label{evf} D=r_s\,\frac{\partial}{\partial r_s}+l\,\frac{\partial}{\partial l},
	\end{equation}
	from which, Eqs. (\ref{lemma}) and (\ref{lemma3}) yield $D R_+= D\alpha=0$, meaning that both $R_+$ and $\alpha$ are homogeneous of degree zero. It is then straightforward to show that
	\begin{equation}
	\label{homfh}D \mathcal{F}_H=0,
	\end{equation}
	and this way, we can infer that $\mathcal{F}_H(\lambda r_s, \lambda l)=\mathcal{F}_H(r_s, l)$.
	The Pfaffian form of the thermodynamics for the Hayward black hole is
	\begin{equation}
	\label{a1}\delta Q_{{\rm rev}}\equiv {\rm d}r_s-\mathcal{F}_H\,{\rm d}l,
	\end{equation}
	where $r_s$ is now identified with the internal energy and $\mathcal{F}_H\,{\rm d}l$ is standard work term. Clearly, the Pfaffian form (\ref{a1}) is homogeneous of degree one, and the Carathéodory's postulate provides
	\begin{equation}
	\label{a2}\delta Q_{{\rm rev}}=\mathcal{T}\,{\rm d}\mathcal{S},
	\end{equation}
	in which $\mathcal{T}$ is the temperature and $\mathcal{S}$ stand for the metrical entropy. 
	One could, therefore, use the homogeneity to connect the Carathéodory and Gibbs approaches in order to justify the thermodynamic construction postulated and demonstrated by Belgiorno \cite{Belgiorno2002,Belgiorno:2002iv}.}




\section{Summary}\label{Summary}

The Hayward black hole corresponds to a regular version of the Schwarzschild black hole, whose deviation is associated with the Hayward's parameter $l$, {obtained from Einstein gravity being coupled to nonlinear electrodynamics}. Hence, for $0\leq l<\sqrt{2}l_0$, with $l_0\equiv \sqrt{2/27} r_s\approx 0.272 r_s$ and $r_s$ as the  classical Schwarzschild radius, the discriminant of the corresponding cubic polynomial associated with the lapse function is positive, and there are two positive roots giving rise to a regular black hole. Here, the Schwarzschild and the extreme black holes correspond respectively to $l=0$ and $l=\sqrt{2} l_0$. 

Considering the above background, in this work, the Bekenstein's theory of entropy was then applied to a thermodynamic pair ($r_s, l$).  Since this definition involves the event horizon $r_+$, we extensively discussed the horizons in terms of both of the aforementioned parameters in Subsect. \ref{roots22}. Obviously, the introduction of the Hayward's parameter makes it necessary to include the generalized Hayward force $\mathcal{F}_H$, as the canonical variable conjugated to it. In this way, we can construct the first and the second laws of thermodynamics, which, if combined together for reversible processes, provide
$$
\mathcal{T}\,{\rm d}\mathcal{S}= {\rm d}r_s-\mathcal{F}_H\,{\rm d}l.
$$

{It is important to note that in our thermodynamic construction, all non-extremal states occur for $\mathcal{T}>0$, while the extremal boundary is characterized by $\mathcal{T}=0$.}

In a similar fashion, we could construct the Hayward's free energy $\Xi$, by making use of the usual thermodynamic potentials, such as enthalpy $\mathcal{H}$, Helmholtz free energy $\mathcal{A}$, and Gibbs free energy $\mathcal{G}$, and it was shown that the minimum and maximum values of $\Xi$ correspond respectively to the Schwarzschild and the extreme Hayward black holes.

The analysis of the curves in the $\mathcal{T}-\mathcal{S}$ diagram, highlights the fact that any change satisfying $\Delta \mathcal{S}\geq 0$, necessarily implies $\Delta l\leq 0$. This way, the Schwarzschild black hole is obtained in the last possible final state, by varying $l$. One could consider this as a confirmation of the third law, in the sense that it is not possible to bring the system to $\mathcal{T} = 0$ in a finite number of steps. This is because by fixing $r_s$, necessarily, $l$ and the corresponding area will grow.

{Lastly, We  studied the heat capacity $\mathcal{C}_l$ by letting $l$ to be a constant and, this way, we brought up the existence of a second order phase transition at $l=l_{pt}\approx 1.22 \,l_0$. This phase transition is isomorphic to the other specific heats, calculated for the black holes, based on their diverse characteristics. This fact also helped us showing that, like the Schwarzschild black hole, by radiating energy, the Hayward black hole warms up for $0\leq l<l_{pt}$, and cools down for $l_{pt}<l\leq \sqrt{2} l_0$.}

{We should note that, the present construction must be refined for the case of a rotating black hole, for which, the Belgiorno approach can be considered to implement black hole thermodynamics through an appropriate quasi-homogeneous Pfiaffan approach \cite{Belgiorno:2002iw}.}

We close our discussion at this point and leave those aspects which were not discussed here, to a forthcoming investigation where we will also provide more profound discussions on the third law \cite{Belgiorno:2002pm,Belgiorno_2003a,Belgiorno_2003b}.

\vspace{1.5mm}


\begin{acknowledgements}
	The authors appreciate important discussions and conversations with M. Fathi, N. Cruz, S. Lepe and A. Pizarro. J.R.V. was partially supported by Centro de Astrof\'isica de Valpara\'iso.
\end{acknowledgements}

\bibliographystyle{ieeetr} 
\bibliography{Biblio_v1.bib}

\begin{thebibliography}{10}

\bibitem{Akiyama:2019cqa}
K.~Akiyama {\em et~al.}, ``{First M87 event horizon telescope results. I. The
  shadow of the supermassive black hole},'' {\em Astrophys. J.}, vol.~875,
  no.~1, p.~L1, 2019.

\bibitem{Schwarzschild:1916uq}
K.~Schwarzschild, ``Über das gravitationsfeld eines massenpunktes nach der
  einsteinschen theorie,'' {\em Berliner Sitzungsbesichte (Phys. Math.
  Klasse)}, vol.~1916, pp.~189--196, 1916.

\bibitem{kottler}
F.~Kottler, ``Über die physikalischen grundlagen der einsteinschen
  relativitätstheorie,'' {\em Ann. d. Physik}, vol.~56, no.~14, pp.~401--462,
  1918.

\bibitem{reissner}
H.~Reissner, ``Über die eigengravitation des elektrischen feldes nach der
  einsteinschen theorie,'' {\em Ann. d. Physik}, vol.~355, no.~9, pp.~106--120,
  1916.

\bibitem{nordstrom}
G.~{Nordstr{\"o}m}, ``Een en ander over de energie van het zwaarte krachtsveld
  volgens de theorie van einstein,'' {\em Proc. Kon. Ned. Akad. Wet.}, vol.~20,
  pp.~1238--1245, Jan. 1918.

\bibitem{Kerr:1963ud}
R.~P. Kerr, ``{Gravitational field of a spinning mass as an example of
  algebraically special metrics},'' {\em Phys. Rev. Lett.}, vol.~11,
  pp.~237--238, 1963.

\bibitem{Newman:1965my}
E.~T. Newman, R.~Couch, K.~Chinnapared, A.~Exton, A.~Prakash, and R.~Torrence,
  ``{Metric of a rotating, charged Mass},'' {\em J. Math. Phys.}, vol.~6,
  pp.~918--919, 1965.

\bibitem{Gibbons:1977mu}
G.~W. Gibbons and S.~W. Hawking, ``{Cosmological event horizons,
  thermodynamics, and particle creation},'' {\em Phys. Rev.}, vol.~D15,
  pp.~2738--2751, 1977.

\bibitem{bardeen}
J.~M. Bardeen, ``Non-singular general-relativistic gravitational collapse,''
  {\em Proc. Int. Conf. GR5, Tbilisi, USSR}, p.~174, 1968.

\bibitem{Hayward06}
S.~A. Hayward, ``Formation and evaporation of nonsingular black holes,'' {\em
  Phys. Rev. Lett.}, vol.~96, p.~031103, Jan 2006.

\bibitem{Dymnikova:1992ux}
I.~Dymnikova, ``{Vacuum nonsingular black hole},'' {\em Gen. Rel. Grav.},
  vol.~24, pp.~235--242, 1992.

\bibitem{Dymnikova:2003vt}
I.~Dymnikova, ``{Spherically symmetric space-time with the regular de Sitter
  center},'' {\em Int. J. Mod. Phys. D}, vol.~12, pp.~1015--1034, 2003.

\bibitem{AyonBeato:1998ub}
E.~Ay\'on-Beato and A.~Garc\'ia, ``{Regular black hole in general relativity
  coupled to nonlinear electrodynamics},'' {\em Phys. Rev. Lett.}, vol.~80,
  pp.~5056--5059, 1998.

\bibitem{AyonBeato:1999ec}
E.~Ay\'on-Beato and A.~Garc\'ia, ``{Nonsingular charged black hole solution for
  nonlinear source},'' {\em Gen. Rel. Grav.}, vol.~31, pp.~629--633, 1999.

\bibitem{AyonBeato:1999rg}
E.~Ay\'on-Beato and A.~Garc\'ia, ``{New regular black hole solution from
  nonlinear electrodynamics},'' {\em Phys. Lett. B}, vol.~464, p.~25, 1999.

\bibitem{AyonBeato:2000zs}
E.~Ay\'on-Beato and A.~Garc\'ia, ``{The Bardeen model as a nonlinear magnetic
  monopole},'' {\em Phys. Lett. B}, vol.~493, pp.~149--152, 2000.

\bibitem{AyonBeato:2004ih}
E.~Ay\'on-Beato and A.~Garc\'ia, ``{Four parametric regular black hole
  solution},'' {\em Gen. Rel. Grav.}, vol.~37, p.~635, 2005.

\bibitem{Mars_1996}
M.~Mars, M.~M. Mart{\'{\i}}n-Prats, and J.~M.~M. Senovilla, ``{Models of
  regular Schwarzschild black holes satisfying weak energy conditions},'' {\em
  Class. Quantum Grav.}, vol.~13, pp.~L51--L58, may 1996.

\bibitem{Borde:1996df}
A.~Borde, ``{Regular black holes and topology change},'' {\em Phys. Rev. D},
  vol.~55, pp.~7615--7617, 1997.

\bibitem{Mbonye:2005im}
M.~R. Mbonye and D.~Kazanas, ``{A Non-singular black hole model as a possible
  end-product of gravitational collapse},'' {\em Phys. Rev. D}, vol.~72,
  p.~024016, 2005.

\bibitem{Bronnikov:2005gm}
K.~Bronnikov and J.~Fabris, ``{Regular phantom black holes},'' {\em Phys. Rev.
  Lett.}, vol.~96, p.~251101, 2006.

\bibitem{Berej:2006cc}
W.~Berej, J.~Matyjasek, D.~Tryniecki, and M.~Woronowicz, ``{Regular black holes
  in quadratic gravity},'' {\em Gen. Rel. Grav.}, vol.~38, pp.~885--906, 2006.

\bibitem{Bambi:2013ufa}
C.~Bambi and L.~Modesto, ``{Rotating regular black holes},'' {\em Phys. Lett.
  B}, vol.~721, pp.~329--334, 2013.

\bibitem{Balart:2014cga}
L.~Balart and E.~C. Vagenas, ``{Regular black holes with a nonlinear
  electrodynamics source},'' {\em Phys. Rev. D}, vol.~90, no.~12, p.~124045,
  2014.

\bibitem{Balart:2016zrd}
L.~Balart and F.~Peña, ``{Regular charged black holes, quasilocal energy and
  energy conditions},'' {\em Int. J. Mod. Phys. D}, vol.~25, no.~06,
  p.~1650072, 2016.

\bibitem{sert}
O.~Sert, ``{Regular black hole solutions of the non-minimally coupled $Y(R)F^2$
  gravity},'' {\em J. Math. Phys.}, vol.~57, no.~3, p.~032501, 2016.

\bibitem{Stuchlik:2014qja}
Z.~e. Stuchlík and J.~Schee, ``{Circular geodesic of Bardeen and
  Ay\'on--Beato--Garc\'ia regular black-hole and no-horizon spacetimes},'' {\em
  Int. J. Mod. Phys. D}, vol.~24, no.~02, p.~1550020, 2014.

\bibitem{Abbas:2014oua}
G.~Abbas and U.~Sabiullah, ``{Geodesic study of regular Hayward black hole},''
  {\em Astrophys. Space Sci.}, vol.~352, pp.~769--774, 2014.

\bibitem{Zhao:2017cwk}
S.-S. Zhao and Y.~Xie, ``{Strong deflection gravitational lensing by a modified
  Hayward black hole},'' {\em Eur. Phys. J. C}, vol.~77, no.~5, p.~272, 2017.

\bibitem{chiba17}
T.~Chiba and M.~Kimura, ``{A note on geodesics in the Hayward metric},'' {\em
  Prog. Theor. Exp. Phys.}, vol.~2017, 04 2017.
\newblock 043E01.

\bibitem{Fernando}
S.~Fernando and J.~Correa, ``{Quasinormal Modes of Bardeen Black Hole: Scalar
  Perturbations},'' {\em Phys. Rev. D}, vol.~86, p.~064039, 2012.

\bibitem{Flachi:2012nv}
A.~Flachi and J.~P. Lemos, ``{Quasinormal modes of regular black holes},'' {\em
  Phys. Rev. D}, vol.~87, no.~2, p.~024034, 2013.

\bibitem{Lin:2013ofa}
K.~Lin, J.~Li, and S.~Yang, ``{Quasinormal modes of Hayward regular black
  hole},'' {\em Int. J. Theor. Phys.}, vol.~52, pp.~3771--3778, 2013.

\bibitem{Saleh:2018hba}
M.~Saleh, B.~B. Thomas, and T.~C. Kofane, ``{Quasinormal modes of gravitational
  perturbation around regular Bardeen black hole surrounded by quintessence},''
  {\em Eur. Phys. J. C}, vol.~78, no.~4, p.~325, 2018.

\bibitem{Mehdipour:2016vxh}
S.~Mehdipour and M.~Ahmadi, ``{Black hole remnants in Hayward solutions and
  noncommutative effects},'' {\em Nucl. Phys. B}, vol.~926, pp.~49--69, 2018.

\bibitem{Huang:2014nka}
H.~Huang, P.~Liao, J.~Chen, and Y.~Wang, ``{Absorption and scattering cross
  section of regular black holes},'' {\em J. Grav.}, vol.~2014, p.~231727,
  2014.

\bibitem{Saleh18}
M.~Saleh, B.~B. Thomas, and T.~C. Kofane, ``{Thermodynamics phase transition of
  regular Hayward black hole surrounded by quintessence},'' {\em Int. J. Theor.
  Phys.}, vol.~57, no.~9, pp.~2640--2647, 2018.

\bibitem{Rodrigue20}
K.~K.~J. Rodrigue, M.~Saleh, B.~B. Thomas, and K.~T. Crepin, ``Thermodynamic
  phase transition and global stability of the regular hayward black hole
  surrounded by quintessence,'' {\em Mod. Phys. Lett. A}, vol.~0, no.~0,
  p.~2050129, 2020.

\bibitem{Fan:2016hvf}
Z.-Y. Fan and X.~Wang, ``{Construction of Regular Black Holes in General
  Relativity},'' {\em Phys. Rev. D}, vol.~94, no.~12, p.~124027, 2016.

\bibitem{Fan:2016rih}
Z.-Y. Fan, ``{Critical phenomena of regular black holes in anti-de Sitter
  space-time},'' {\em Eur. Phys. J. C}, vol.~77, no.~4, p.~266, 2017.

\bibitem{Bronnikov:2000vy}
K.~A. Bronnikov, ``{Regular magnetic black holes and monopoles from nonlinear
  electrodynamics},'' {\em Phys. Rev. D}, vol.~63, p.~044005, 2001.

\bibitem{Bronnikov:2017tnz}
K.~A. Bronnikov, ``{Comment on \textquotedblleft{}Construction of regular black
  holes in general relativity\textquotedblright{}},'' {\em Phys. Rev. D},
  vol.~96, no.~12, p.~128501, 2017.

\bibitem{Toshmatov:2018cks}
B.~Toshmatov, Z.~Stuchl\'\i{}k, and B.~Ahmedov, ``{Comment on
  \textquotedblleft{}Construction of regular black holes in general
  relativity\textquotedblright{}},'' {\em Phys. Rev. D}, vol.~98, no.~2,
  p.~028501, 2018.

\bibitem{lang02}
S.~Lang, {\em Algebra}.
\newblock New York, NY: Springer, 2002.

\bibitem{Cruz:2004ts}
N.~Cruz, M.~Olivares, and J.~R. Villanueva, ``{The geodesic structure of the
  Schwarzschild anti-de Sitter black hole},'' {\em Class. Quantum Grav.},
  vol.~22, pp.~1167--1190, 2005.

\bibitem{fkov}
M.~Fathi, M.~Kariminezhaddahka, M.~Olivares, and J.~R. Villanueva, ``{Motion of
  massive particles around a charged Weyl black hole and the geodetic
  precession of orbiting gyroscopes},'' {\em Eur. Phys. J. C}, vol.~80, no.~5,
  pp.~1--20, 2020.

\bibitem{Bekenstein:1972}
J.~D. Bekenstein, ``Black holes and the second law,'' {\em Lettere al Nuovo
  Cimento (1971-1985)}, vol.~4, pp.~737--740, Aug 1972.

\bibitem{Bekenstein:1973}
J.~D. Bekenstein, ``Black holes and entropy,'' {\em Phys. Rev. D}, vol.~7,
  pp.~2333--2346, Apr 1973.

\bibitem{Bekenstein:1974}
J.~D. Bekenstein, ``Generalized second law of thermodynamics in black-hole
  physics,'' {\em Phys. Rev. D}, vol.~9, pp.~3292--3300, Jun 1974.

\bibitem{Bekenstein:1975}
J.~D. Bekenstein, ``Statistical black-hole thermodynamics,'' {\em Phys. Rev.
  D}, vol.~12, pp.~3077--3085, Nov 1975.

\bibitem{Bardeen:1973gs}
J.~M. Bardeen, B.~Carter, and S.~Hawking, ``{The Four laws of black hole
  mechanics},'' {\em Commun. Math. Phys.}, vol.~31, pp.~161--170, 1973.

\bibitem{Hawking:1974sw}
S.~Hawking, ``{Particle Creation by Black Holes},'' {\em Commun. Math. Phys.},
  vol.~43, pp.~199--220, 1975.
\newblock [Erratum: Commun.Math.Phys. 46, 206 (1976)].

\bibitem{fmv20}
M.~Fathi, M.~Molina, and J.~R. Villanueva, ``{Adiabatic evolution of Hayward
  black hole},'' 2021.

\bibitem{Belgiorno2002}
F.~{Belgiorno}, ``{Homogeneity as a bridge between Carath{\'e}odory and
  Gibbs},'' {\em arXiv e-prints}, pp.~math--ph/0210011, Oct. 2002.

\bibitem{Belgiorno:2002iv}
F.~Belgiorno, ``{Black hole thermodynamics in Caratheodory's approach},'' {\em
  Phys. Lett. A}, vol.~312, pp.~324--330, 2003.

\bibitem{Misner:1974qy}
C.~W. Misner, K.~Thorne, and J.~Wheeler, {\em {Gravitation}}.
\newblock San Francisco: W. H. Freeman, 1973.

\bibitem{reichl2009modern}
L.~Reichl, {\em A modern course in statistical physics}.
\newblock Physics textbook, Wiley, 2009.

\bibitem{Belgiorno:2002pm}
F.~Belgiorno and M.~Martellini, ``{Black holes and the third law of
  thermodynamics},'' {\em Int. J. Mod. Phys.}, vol.~D13, pp.~739--770, 2004.

\bibitem{Belgiorno_2003a}
F.~{Belgiorno}, ``{Notes on the third law of thermodynamics: I},'' {\em Journal
  of Physics A Mathematical General}, vol.~36, pp.~8165--8193, Aug. 2003.

\bibitem{Belgiorno_2003b}
F.~Belgiorno, ``Notes on the third law of thermodynamics: {II},'' {\em Journal
  of Physics A: Mathematical and General}, vol.~36, pp.~8195--8221, jul 2003.

\bibitem{Kastor:2009wy}
D.~Kastor, S.~Ray, and J.~Traschen, ``{Enthalpy and the mechanics of AdS black
  holes},'' {\em Class. Quant. Grav.}, vol.~26, p.~195011, 2009.

\bibitem{davies77}
P.~C.~W. Davies, ``The thermodynamic theory of black holes,'' {\em Proceedings
  of the Royal Society of London. Series A, Mathematical and Physical
  Sciences}, vol.~353, no.~1675, pp.~499--521, 1977.

\bibitem{ruppeiner08}
G.~Ruppeiner, ``{Thermodynamic curvature and phase transitions in Kerr-Newman
  black holes},'' {\em Phys. Rev. D}, vol.~78, p.~024016, Jul 2008.

\bibitem{Belgiorno:2002iw}
F.~Belgiorno, ``{Quasihomogeneous thermodynamics and black holes},'' {\em J.
  Math. Phys.}, vol.~44, pp.~1089--1128, 2003.

\end{thebibliography}

\end{document}